\documentclass[10pt]{article}
\usepackage{graphicx}
\usepackage{amssymb}
\usepackage{amsmath}
\usepackage{epsf}
\usepackage{cite}
\usepackage{color}

\usepackage[labelfont=bf,labelsep=period,justification=raggedright]{caption}

\begin{document}

\begin{flushleft}
{\Large\textbf{A seed-diffusion model for tropical tree diversity patterns}}\\
A. Derzsi$^{1}$, Z. N\'{e}da$^{1,2}$ \\
{\small\textbf{1 Department of Theoretical and Computational Physics, Babe\c{s}-Bolyai University, Kog\u{a}lniceanu street 1, RO-400080, Cluj-Napoca, Romania}}\\
{\small\textbf{2 Interdisciplinary Computer Modeling Group, Hungarian University Federation of Cluj, Baba Novac street 23/2, RO-400080, Cluj-Napoca, Romania}}\\
\end{flushleft}

\section*{Abstract}
Diversity patterns of tree species in a tropical forest community are approached by a simple lattice model and investigated by Monte Carlo simulations using a
backtracking method.  Our spatially explicit neutral model is based on a simple statistical physics process, namely the diffusion of seeds. The model has three parameters:
the speciation rate, the size of the meta-community in which the studied tree-community is embedded, and the average surviving time of the seeds.
By extensive computer simulations we aim the reproduction of relevant statistical measures derived from the experimental data of the Barro Colorado Island tree census in year 1995.
The first two parameters of the model are fixed to known values, characteristic of the studied community, thus obtaining a model with only one freely adjustable parameter.
As a result of this, the average number of species in the considered territory, the relative species abundance distribution, the species-area relationship and the
spatial auto-correlation function of the individuals in abundant species are simultaneously fitted with only one parameter which is the average surviving time of the seeds.

\section{Introduction}

The endeavor to explain the observed statistical patterns of species abundance, distribution and diversity in ecological communities and to reveal the general mechanisms responsible for these patterns motivated a great number of scientific works in theoretical ecology and statistical physics in the last decades. Initial attempts in this sense consisted in fitting statistical distributions to species abundance data of various samples without paying attention to the underlying ecological processes \cite{FisherEtAl_1943, Preston_1948}. Theories counting the presence of some governing forces in the development of the structure and biodiversity of ecological communities have been also elaborated. The traditional theory of niche differentiation attributes this guiding force to competition \cite{Hutchinson_1957, Levins_1968, MacArthur_1970, Tilman_1982}. According to this approach, each species in a given community is the best competitor in its own ecological niche, therefore no two or more species can coexist in the same niche for a long time. In the view of the niche-assembly theory, species can live together in a community only when they exhibit differences from one another in the use of the available resources. This approach proved to be successful in describing the distribution and abundance of species in various environments \cite{Silvertown_2004}, but evidently, it has some difficulties in explaining the richness of species often observed in communities comprising ecologically similar species, such as tropical forests. The neutral theory \cite{Hubbell_1997, Hubbell_2001}, introduced as an alternative to the niche theory, proposed a striking aspect: it propagates the functional equivalence of trophically similar species and attributes the key role to randomness as a driving force in the development of the observed ecological patterns.

Since its publication, the neutral theory has received much attention. This new approach provided the first models to explain species abundance based on fundamental biological processes of species extinction, immigration or speciation \cite{Hubbell_2001, Hubbell_2006, Bell_2001}. The striking similarities with classical statistical physics models mad it very attractive for the physics community \cite{VolkovEtAl_2003,Pigolotti_2004,ZillioEtAl_2005,VolkovEtAl_2007,Etienne_2007}.  Despite their amazing simplicity, neutral models produced accurate predictions for some taxa \cite{Hubbell_2001, VolkovEtAl_2003, VolkovEtAl_2007}. However, the assumption of functional equivalence of individuals regardless of species in an ecological community has generated serious debates among ecologists \cite{McGill_2003, Hubbell_2005, AlonsoEtAl_2006, Gevin_2006}. In these days, the neutral approach is more and more received as a null model of community dynamics and attempts are made to bridge it to the niche-assembly theory \cite{Gravel_2006, Leibold_McPeek_2006, ChuEtAl_2007}.

The neutral theory of biodiversity is considered analogous to the neutral theory of molecular evolution in population genetics \cite{Kimura_1983}. In fact, the neutral theory in ecology traces back to the theory of island biogeography \cite{MacArthur_Wilson_1967} and extends it by taking into account speciation and the number of coexisting species as well. In this simple approach space is modeled implicitly: a local community is considered where speciation balances extinction by receiving migrants from a large external source pool (metacommunity) and its dynamics is realized at the level of individuals. Besides the functional equivalence of all individuals of all the species in a local community, the neutral theory also assumes a constant size of the community predicting zero-sum multinomial distribution for the species abundance distribution.

Application of neutral models to describe the relative species abundance (RSA) distribution of tropical rain forest communities turned out to be especially resultful. In Hubbell's model \cite{Hubbell_2001, VolkovEtAl_2003}, the ecological drift, without any additional mechanisms, is sufficient to produce patterns of species diversity and abundance in accordance with those observed in nature. In the past decade, a series of neutral models based on Hubbell's simple spatially implicit model have been developed \cite{VolkovEtAl_2007,Dewdney_2000, Enquist_2002,Chave_2004,Pigolotti_2004,Etienne_2007}. Recently, models with various spatial structures  \cite{MunozEtAl_2007, Economo_Keitt_2008, Etienne_2009} and spatially explicit models  \cite{Chave_Leigh_2002, ZillioEtAl_2005, Rosindell_Cornell_2007, HorvatEtAl_2010, ODwyer_Green_2010} have been introduced. Such neutral models present the metacommunity in a more realistic way and are able to capture to a larger extent the effect of individual birth, death, dispersal and speciation processes. Spatially explicit neutral models performed well in the prediction of the observed species-area relationships (SAR) in tropical tree communities. While the simplicity of spatially implicit (mean-field type) models facilitates their analytical treatment, the more complex spatially explicit models exhibit limited analytical tractability.

In the present work a spatially explicit neutral model for tropical tree diversity patterns is introduced and studied. There are a great number of ecological patterns worthy of note but reproduction of all these patterns simultaneously exceeds the ability of current modeling approaches. However, the importance of fitting multiple patterns at the same time is undisputed as this could better reveal the correct contribution of ecological processes to shaping real ecological communities. The motivation of this paper comes thus from the endeavor to construct a simple spatially explicit neutral model that, besides the most studied RSA and SAR measures, can effectively capture another experimentally verifiable macroecological measure, the spatial distribution of individuals of a species, characterized by e.g. the spatial autocorrelation function (SAF).

In the literature there are spatial models that can reproduce the observed species abundance distribution together with the species-area scaling, however, the applicability of these models to reproduce in the same time the observed spatial distribution of individuals of a species generally is not investigated or attempts concerning this purpose failed. Our previously developed spatially explicit neutral model \cite{HorvatEtAl_2010} also shows this limited efficiency.  While it is successful in reproducing the measured RSA and SAR in tropical tree communities, it produces exponential-type decay for the spatial autocorrelation function (SAF) of abundant species instead of the expected power-law type decay.

Our model is based on the diffusion of seeds \cite{Chave_comparing_2002}, the diffusion assumed to be a simple random walk in the two dimensional space. The maximal surviving time of seeds defines the scale of the dispersal kernel, which describes the probability for an individual of a species to occupy locations at different distances in the ecosystem. For dispersal rule, one can find various assumptions in the literature \cite{Chave_Leigh_2002,ZillioEtAl_2005, Rosindell_Cornell_2007, Chave_comparing_2002, Durrett_Levin_1996}. As already pointed out in \cite{Chave_comparing_2002}, the choice of the dispersal distance is crucial in seed diffusion models as it strongly influences the shape of the obtained species-area curves: on a log-log scale, models with nearest-neighbor dispersal result in convex curves, while models with global dispersal produce concave curves.

The approach considered  here is based on a three-parameter model, one of these parameters being a freely adjustable variable: the average surviving time of seeds (common for all species in the local community). The other two parameters are set to realistic values corresponding to the investigated system. However, mention must be made of the incertitude of the speciation rate: since its value is difficult to measure in practice and in neutral models one can find values for this parameter taken from a wide range, the value of the speciation rate set here may also seem speculative. The size of the metacommunity is fixed according to the area of an extended region clearly delimited by natural borders (e.g. a region delimited by see, ridge of mountains), surrounding the studied region.
In real communities, the dispersal of seeds can be influenced or restricted by properties of the natural environment itself. Delimiting thus the size of the metacommunity, the model is adaptable to a spatial region where the dispersal process of seeds can be described properly by assuming the same dispersal rules. In this manner the model essentially provides a simple one-parameter fit for several macroecological measures of the investigated tropical tree community: RSA, SAR, species number and SAF.

The model is applied here to reproduce the ecological patterns obtained for the Barro Colorado Island (BCI) tropical tree census \cite{Condit_1998,Hubbel_Condit_Foster_2005}, the oldest and most studied moist forest plot. A number of individual trees (corresponding to the number of trees in the studied region of the BCI plot) are positioned on a uniform grid, representing the survey area in the model. Computer simulation of the model is realized by using the backtracking coalescence technique, described in \cite{Rosindell_Wong_Etienne_2008} and applied in \cite{Rosindell_Cornell_2007,Etienne_Olff_2004,Pigolotti_Cencini_2009}, a powerful simulation method for neutral models in ecology. This method, consisting of tracing lineages of all individuals originating in the survey area backward in time, until they speciate or coalesce, proved to be efficient for large and complex spatial structures (even for the case of infinite landscapes). In the present study, around the survey area, a region corresponding to the area of the whole BCI is involved in the calculations, representing the metacommunity, and a speciation rate for tropical tree communities is assumed in the model. Computer simulations of the model focus on the reproduction of the number of species, the relative species abundance distribution, the species-area scaling and the spatial distribution of individuals of abundant species in the BCI plot, by investigating the effect of the average surviving time of the seeds.
Since the model has only one freely adjustable parameter, our goal to reproduce simultaneously all the mentioned macro-ecological measures seems quite ambitious. In such case one should not expect to obtain better fits for all these measures than the fits provided by different models focusing on specified ecological measures, considering only one or just a few of them from the whole set of relevant characteristics. Fitting only one or two of these measures can definitely result in considerably better fits, however, neglecting completely the reproduction of other important properties. In our view a successful model should provide a description for all the measurable macroecological patterns, and we consider thus that obtaining acceptable qualitative fits, simultaneously valid for the relevant measures, is more important than an excellent fit for one or just a few measures.

\section{Experimental data}

As experimental data, the BCI tropical tree census data is used \cite{Condit_1998,Hubbel_Condit_Foster_2005}. Based on this, the relevant macroecological measures are defined and compared with simulation outcomes.

The BCI, administrated by the Smithsonian Tropical Research Institute, a biological reserve from 1923, is located in the Gatun Lake, Panama. The island is covered with moist forest. Since 1980, a 50 $ha$ (1000 m $\times$ 500 m) region is regularly mapped, in each census all free standing woody stems with a stem diameter at least 1 cm at breast height are identified, tagged and recorded. The whole dataset contains information on about 240000 stems of more than 300 tree and shrub species, based on this, accurate statistics can be made. Census data is publicly available for several years.

In this work census data from the year 1995 is used. For this year, 243541 individuals of 302 species with at least 1 cm stem diameter can be found in the dataset. Analyzing the density of individuals in the 50 ha plot, a less populated region can be identified in the western part of it, corresponding to a swampland in the studied area (this pattern is observable independently of census year). To ensure homogeneity when calculating the ecological characteristics, only a subregion of the entire plot, limited to the eastern 500 m $\times$ 500 m section, is considered. This 25 $ha$ region contains 112543 trees and shrubs of 273 species.

\section{Model}

We investigate a local community in a spatial region of size $L \times L$ embedded in a much larger territory (e.g. island, continent, infinite land). Our model follows the idea formulated in \cite{Chave_comparing_2002} with the notable difference in the determination of the used dispersal kernel: instead of postulating a dispersal kernel without arguing its relation to ecological processes, here it is derived from a real physical process based on the diffusion of seeds. Following separately the random motion of each seed is a computationally demanding task due to the large number of individuals and seeds present in the considered community, therefore a statistical description of the spatial distribution of seeds is considered.

Our basic hypothesis is that from time to time one individual dies and its place is taken by a new individual originating from a randomly chosen seed found in its neighborhood. It is assumed that (i) each individual emits the same number of seeds at each time-step, which randomly diffuse in the 2D space by a simple unbiased random walk; (ii) the seeds survive $W$ time steps, after which they die and become unable to grow; (iii) all the seeds (even those belonging to different species) have the same diffusion constant and the same surviving period $W$.

\begin{figure}[ht]
 \centering
  \begin{center}
    \includegraphics[scale=0.3]{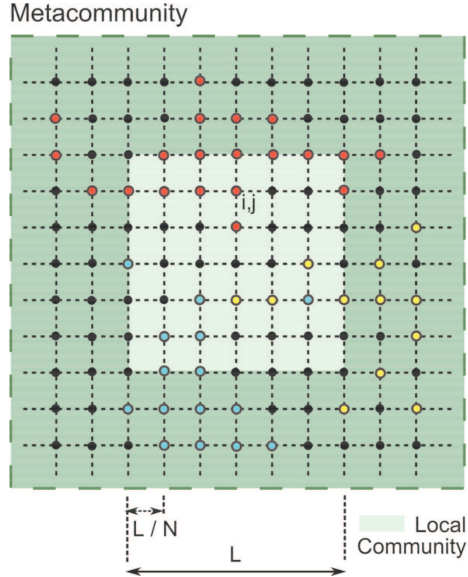}
 \end{center}
\caption{Schematic view of the modeled area. The light-coloured square region represents the studied area of size $L \times L$ embedded in a much larger territory. Individuals are placed to a uniform square grid ($N \times N$), labeled by grid coordinates $(i,j)$. The black circles represents individuals in traced ''active'' lineages, color circles stand for lineages finalized by speciation.}
\label{fig:model}
\end{figure}

Individuals of the studied region with area $L \times L$ are placed on a uniform grid of size $N \times N$. The schematic picture of the studied region is drawn in Fig. \ref{fig:model}. The individuals are labeled by mesh coordinates $(i,j)$. In the neighborhood of each individual there can be seeds originating from other individuals, and we denote by $Q_{ij}(k,l)$  the number of seeds in the neighborhood of individual at grid position $(i,j)$ originating from the individual at grid position $(k,l)$.

A simple two-dimensional unbiased random walk is assumed for all the seeds. The $Q_{ij}(k,l)$ numbers are computed for each pair of individuals, according to the basic statistical properties of simple random walks. It is assumed that in unit time a seed can make a unit step defined as the grid constant $L/N$, randomly moving to an individual in the neighborhood of its current position in a similar manner as the simple unbiased 2D random walker. This defines the time-unit in the simulations. In such case the probability $P_n(x,y)$ to find the random walker at position $(x,y)$ after $n$ number of steps when started from coordinates $(0,0)$ writes as:
\begin{equation}
P_n(x,y)=\frac{1}{\pi n}\exp\left(-\frac{x^2+y^2}{n}\right).
\end{equation}

Assuming that the seeds can survive $W$ time steps, the number of seeds originating from $(0,0)$ that will be in position $(x,y)$ can be calculated as:
%Q_{x,y}(0,0) \approx c K_{0,0} \int_{1}^{W} \! \frac{1}{\pi z}\exp\left(-\frac{r^2}{z}\right) \, \mathrm{d}z,
\begin{equation}
Q_{x,y}(0,0) \approx c \int_{1}^{W} \! \frac{1}{\pi z}\exp\left(-\frac{r^2}{z}\right) \, \mathrm{d}z,
\end{equation}
where $r=\sqrt{x^2+y^2}$ and $c$ is a constant that quantifies the number of seeds emitted by individuals in unit time. Expressing the integral term in the above equation by incomplete gamma functions, the number of seeds in the neighborhood of the position $(i,j)$ originating from an arbitrary $(k,l)$ position can be computed as
%Q_{i,j}(k,l) \approx c K_{k,l} \left[\Gamma(0,\frac{r^2}{W})-\Gamma(0,r^2)\right],
\begin{equation}
Q_{i,j}(k,l) \approx c \left[\Gamma(0,\frac{r^2}{W})-\Gamma(0,r^2)\right],
\label{eq:Qijkl}
\end{equation}
where $r=\sqrt{(i-k)^2+(j-l)^2}$ and the incomplete gamma function $\Gamma(0,y)$ is defined as:

\begin{equation}
\Gamma(0,y)=\int_y^{\infty} \frac{e^{-t}}{t} dt.
\end{equation}

According to the $Q_{i,j}(k,l)$ quantities one can select the "successful" seed that will generate the new individual in an empty place. Details of this dynamical model are given below.

The model is initialized as follows. Individuals are positioned on a uniform $N \times N$ grid and for each individual a lineage is considered. The $Q_{i,j}(k,l)$ quantities are defined for all the $(i,j), (k,l) \in N \times N$ pairs. Initially all the lineages are traced using the coalescence technique  \cite{Rosindell_Wong_Etienne_2008}. The dynamics of the model is defined by the following rules:
\begin{itemize}
 \item In each step one individual is randomly selected from all the lineages being traced at that moment and its $(i,j)$ grid coordinates are determined. The selected individual is the one that was most recently born. The origin of this individual has to be defined.
 \item With probability $q$, speciation happens in the lineage. If speciation takes place, tracing of the lineage is stopped and the selected individual is considered to be the root of a new species.
 \item With probability $1-q$, a parent is defined for the selected individual. The position of the parent is chosen according to the dispersal kernel. Coalescence occurs when two lineages share the same spatial position.
\end{itemize}

\begin{figure}[ht]
 \centering
  \begin{center}
    \includegraphics[scale=0.3]{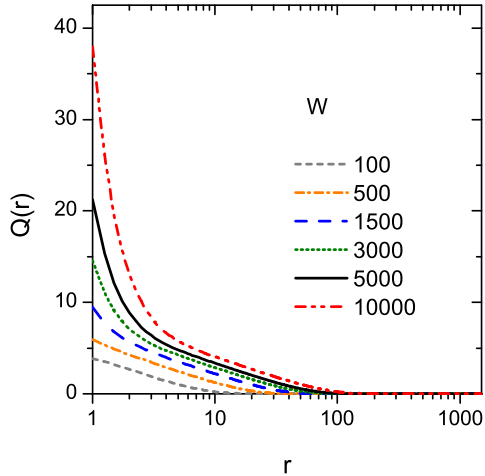}
 \end{center}
\caption{The number of seeds $Q(r)$ in the neighborhood of a given position originating from a location at distance $r$ as a function of the distance calculated for different values of the seed survival time, $W$. $r$ is measured in units of the grid constant $L/N$.}
\label{fig:kernel}
\end{figure}

The parent selection mechanism of the last rule is done by respecting the weights given by the $Q(r)=Q_{i,j}(k,l)$ values (see Fig. \ref{fig:kernel}). Based on these values, a seed originating from an individual at $(k,l)$ is chosen, thus, the individual at $(k,l)$ which emitted the "successful" seed reproduces itself at the position $(i,j)$.

This selection process realizes the simple unbiased two-dimensional diffusion of seeds in the considered area, allowing also seeds originating from distant locations to reach and occupy the vacant position.

The above defined dynamical model can be easily implemented in a computer simulation. Large systems, containing the same number of individuals as a known experimental dataset, can be studied in reasonable computational time.

\section{Computational details}

The model described above is applied here to study the tropical forest community of BCI. In principle, for a fixed community with $N\times N$ individuals the model has three parameters: the lifetime of the seeds $W$, the size of the surrounding metacommunity and the speciation rate, $q$. As experimental data, we use census data for the eastern (square like) half part of the whole sampled territory (25 ha) of BCI for the year 1995. The total number of individuals in this region, $K$, is 112543. In order to have the number of individuals in the model in accordance with the experimental number of individuals, the uniform $N \times N$ square grid is constructed with $N$=335. This grid, representing the survey area from BCI is embedded in the center of a 1130$\times$1130 square grid, representing in a proportional manner the whole island of Barro Colorado (1560 ha). During simulation, tracing of lineages is realized in this larger grid, the position of parents, selected according to the proposed dispersal kernel, can be positions in this larger grid. Simulations have been performed by fixing the value of the speciation rate at a value comparable to the mutation rate of a neutral gene. In agreement with the accepted order of magnitude in the literature \cite{Rosindell_Cornell_2009}, $q=1.0 \times 10^{-5}$ is set as speciation rate and the maximum seed survival period $W$ is varied from 50 to 5000 steps in searching for a $W$ parameter reproducing the experimental value of the total number of species ($S=273$) and the other observed ecological patterns. As already stated in the introduction, our model is essentially a one-parameter model, since fixing the speciation rate and the size of the metacommunity, we are left with only one freely adjustable fitting parameter.

\section{Quantities of interest}

Apart of the total number of individuals and species in the considered local community there are several relevant distributions that should be compared with the census data. For a detailed description of these quantities one could consult our previous work \cite{HorvatEtAl_2010}.

For the spatial distribution of species and individuals two relevant quantities are calculated and compared with experimental results: the Species-Area Relationship (SAR) and the Spatial Autocorrelation Function (SAF) of individuals belonging to the same species.

The SAR is studied by considering larger and larger territories and counting the number of species
present within these areas. Average species numbers are calculated on subregions with the same size,
the SAR curve representing these averages as a function of the considered area

The SAF,  denoted by $C_s(r)$ for species $s$, is constructed in the following way: a uniform mesh (not necessarily the same division as for the mesh of the simulation) is considered on the studied area and in each ($i,j$) cell of this mesh, the number of individuals $K_{i,j}^{(s)}$ from the considered species is determined. The auto-correlation function for a relative coordinate ($p,q$) is calculated as
\newcommand{\avgK}{\ensuremath{\langle K_s \rangle}}
\begin{equation}
  C_{p,q}^{(s)} = \bigl<(K_{i,j}^{(s)} - \avgK) (K_{i+p,j+q}^{(s)} - \avgK) \bigr>_{i,j}
\end{equation}
Here $\avgK$ denotes the average number of individuals of the considered species on the mesh sites: $\avgK = \langle K_{i,j}^{(s)} \rangle_{i,j}$.
The average of the $C_{p,q}^{(s)}$ values for all ($p,q$) mesh coordinates that are inside a ring with radius $r$ and width $\Delta r$ ($r \le \sqrt{p^2 + q^2} \le r + \Delta r$), considering a reasonably small $\Delta r$ value, leads to our definition of SAF:
\begin{equation}
  C_s(r) = \frac{\bigl< C_{p,q}^{(s)} \{ r \le \sqrt{p^2 + q^2} \le r + \Delta r \} \bigr>_{p,q}}{\Delta r}
\end{equation}

Another important quantity that should be also compared with the experimental results is the Relative Species Abundance (RSA) distribution. This distribution characterizes the frequency of species with a given abundance. Different types of representations are generally used to study RSA. The most widespread method to present species abundances is due to F.~W.~Preston \cite{Preston_1962}, who sorted the species of a sample into abundance intervals of consecutively doubling lengths ($[1,2)$, $[2,4)$, $[4,8)$, etc.), and plotted the number of species found within these "octaves". This kind of plotting is motivated by the fact that species abundances can vary over several orders of magnitude and usually there are fewer abundant species than rare ones. This way the presence of large statistical fluctuations at the tail of the curve can be avoided.
A second way of representing the species-abundance distribution is plotting the associated probability density function $\rho(v)$, i.e.  the probability for finding a species with a given $v$ abundance. $\rho(v)$ can be derived from the Preston plot by dividing the number of species in each interval with the length of the interval, and plotting on a double logarithmic plot this quantity versus the mean abundance in the given interval.
A third way of representing the species abundance distribution can be realized by arranging the species in decreasing order by their abundances and plotting the rank of the species versus its abundance (rank-abundance plot) again on a double-logarithmic graph.  This representation is inspired by several abundance studies in sociology and economics, leading to the general Pareto-Zipf distribution.

\section{Results}

Fixing the value of $q=1.0 \times 10^{-5}$, we study the above discussed quantities of interest as a function of $W$, and compare the results with the ones computed from the BCI data.

\begin{figure}[ht]
 \centering
  \begin{center}
    \includegraphics[scale=0.7]{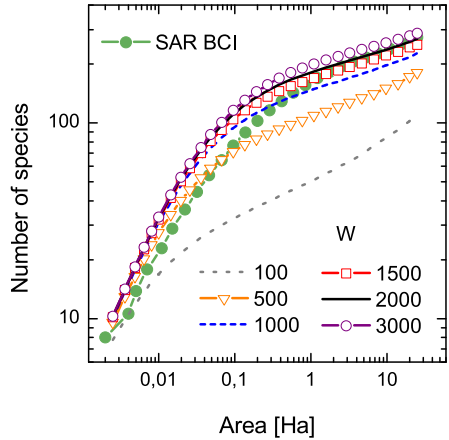}
 \end{center}
\caption{Species-area curves. Simulation results (different lines or lines with open symbols) are presented for various values of the $W$ seed surviving period with fixed speciation rate ($q=1.0 \times 10^{-5}$). The continuous line with filled circles shows the species-area relation (SAR) for the Barro Colorado Island (BCI) census data (census year: 1995).}
\label{fig:sarN100}
\end{figure}

In Fig. \ref{fig:sarN100} results for the SAR curves are presented. Simulations are performed with $W$ values ranging from 100 up to 3000. With increasing $W$ values, the experimental SAR curve is better and better reproduced by our model.

\begin{figure}[ht]
 \centering
  \begin{center}
    \includegraphics[scale=0.7]{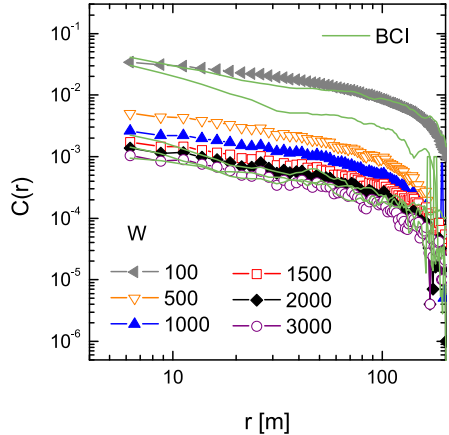}
 \end{center}
\caption{Autocorrelation functions of the most abundant species. Simulation results (lines with symbols - averaged curves) are presented for different values of the $W$ seed surviving period with fixed speciation rate ($q=1.0 \times 10^{-5}$). Results obtained for the five most abundant species in the Barro Colorado Island (BCI) census data are shown with lines (census year: 1995).}
\label{fig:correlationN100}
\end{figure}

In Fig. \ref{fig:correlationN100} the SAF for the most abundant species obtained  with different values of the $W$ parameter are presented. For a visual comparison, the SAF for the $1^{st}$, $2^{nd}$, $3^{rd}$, $5^{th}$ and $10^{th}$ most abundant species in the BCI census, \textit{Hybanthus prunifolius} (17499 individuals), \textit{Faramea occidentalis} (12331 individuals), \textit{Trichilia tuberculata} (8684 individuals), \textit{Alseis blackiana} (4121 individuals) and \textit{Capparis frondosa} (2078 individuals) respectively, are also plotted. Simulation curves are averaged results of 100 runs.  It should be noted that in the simulations, even if the total number of species is the same as in the BCI census, the number of individuals of the most abundant species can show considerable fluctuations, this is the reason why the simulation curves are vertically shifted relative to the BCI results.

\begin{figure}[ht]
 \centering
  \begin{center}
    \includegraphics[scale=1.6]{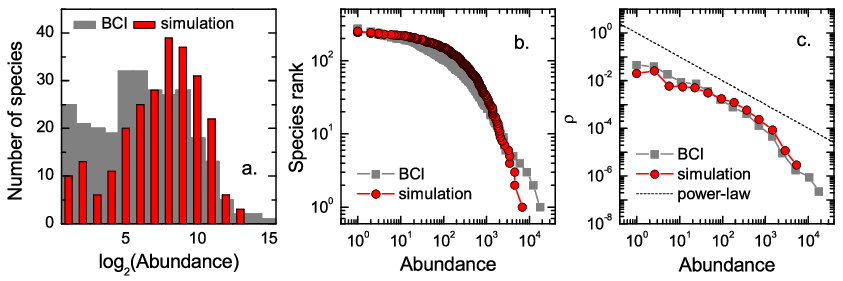}
 \end{center}
\caption{Relative species abundance (RSA). The RSA is plotted in different ways: (a) Preston-plot, (b) rank-abundance plot and (c) probability density function. Simulation results are presented for $W=1700$ steps seed surviving period and $q=1.0 \times 10^{-5}$ speciation rate. Experimental data obtained from the Barro Colorado Island (BCI) dataset is also plotted (census year: 1995). For guiding the eye in panel (c) a power-law with exponent $-1$ is also indicated with a dashed line.}
\label{fig:rsa}
\end{figure}

From the SAR and SAF results one can learn that the spatial distribution of the species and individuals are comparable to the experimental ones for $W\approx1500$ steps. The RSA results for $W=1700$ are presented in Fig. \ref{fig:rsa}. All three accepted representations of the RSA distribution are plotted and the results are compared with those obtained for the BCI census data of year 1995.

We conclude thus that for $W\approx1500$ steps and $q=1.0 \times 10^{-5}$ parameters the model generates distribution of
individuals and species in agreement with the experimental results offered by the BCI dataset. Moreover, the total number of species in the system (local community) is also in agreement with the number of species in the original BCI dataset.

\section{Discussion}

A spatially explicit neutral model was elaborated to reproduce the observed ecological measures for a tropical tree community. Monte Carlo simulations were done on a square-like territory embedded in a larger square-like metacommunity, considering a uniform square lattice topology for tree positions and using the backward simulation technique.

In the model two main processes are responsible for the diversity of species: the dispersal of seeds capable of growth and the speciation, the parameters governing these two processes are $W$ and $q$, respectively.
The dispersal of seeds is interpreted here as collection of individual and independent random walks on the 2D plane. The $W$ parameter controls the maximum possible dispersal distance (surviving time of seeds) and reflects the importance and validity limits of long-range interactions.
The $q$ parameter explicitly specifies the rate of speciation within the considered local community.

The studied area was represented by a uniform grid of $335 \times 335$ positions of individuals, embedded in a larger grid of $1130 \times 1130$ positions. This larger grid models the metacommunity, which in our case is the whole island. The number of trees in the whole island is not known, therefore the size of the larger grid was selected proportionally with the size of the BCI. By fixing the speciation rate to a reasonable $q=1.0 \times 10^{-5}$ order of magnitude, we remain with a one parameter ($W$) model.

Initially a maximum survival distance of 100 steps was assumed, corresponding to a trail of length 150 m. For this initial combination of parameters the experimental SAR is not reproduced, there are large discrepancies especially in the limit of large areas. Simulations yield smaller number of species than expected, which suggests that the intermixture of species in not effective enough.  This means that the limit of long-range interactions (governed by $W$) is farther away than 150 m, which was initially assumed. In Fig. \ref{fig:sarN100} the SAR curves with maximum survival distance between 100 and 3000 steps, corresponding to trail lengths between 150 m and 4491 m are presented. For maximum survival distances above 1500 steps, corresponding to a physical distance of a  2247 m, the SAR curves from the simulations exhibit the typical power-law like behavior of the experimental SAR curves in the limit of large a\-re\-as. Increasing the maximum survival distance facilitates the intermixing of species, resulting in simulation curves more and more close to the experimental one. For $W=1500$ steps, comparison of the simulation outcome and experimental SAR curves shows 22 $\%$ difference of the results.
However, one has to be aware of the SAF of abundant species as well. For the abundant species of the BCI census area, this function shows power-law type decrease for not too large distances. This power-law type decrease was not well-reproduced by the earlier spatially extended models. In Fig. \ref{fig:correlationN100} the variation of the SAF is presented for different survival distances in comparison with experimental SAF curves. Analyzing the simulation results for values of the survival distance $W>1500$ steps, the power-law decrease is reproduced, in agreement with the experimental SAF curves.
Following the third measure of interest, the RSA, as a function of $W$, the $W=1700$ steps case proves to be the most appropriate to reflect the characteristic ecological measures of the BCI plot (Figure \ref{fig:rsa}). For this case, the average difference in the simulation and experimental results is quite high: 53 $\%$. This large difference should be judged by taking into account that we fit simultaneously four relevant ecological measures with only one scalar parameter.

The results obtained for $W=1700$ and $q=1.0 \times 10^{-5}$ are acceptable thus for the ecological measures considered here. Naturally, none of the fits for the relevant ecological measures is perfect, one cannot expect this from a simple one-parameter neutral model. Real tropical tree communities are complex systems, and nowadays it is expected  that neutral models
should be viewed as null models of community dynamics, and niche-assembly elements are also needed to obtain a better description of the systems. The non-perfect description offered by this simple approach confirms this view.

\section{Conclusion}

The statistical patterns observed in a tropical tree community were theoretically approached by a spatially explicit neutral model based on the diffusion of seeds. The considered model has three parameters, the maximum diffusion distance of seeds, the probability of speciation in the system and the size of the metacommunity. Two of these parameters (the size of the metacommunity and the speciation rate) were fixed according to the experimental conditions.
In principle, we remained thus with a simple one-parameter  model and this parameter ($W$) was adjusted in order to reproduce the major ecological measures obtained from the BCI tropical tree census. Considering the same number of individuals in the
model as in the BCI census, we were searching for a proper $W$ value, which would simultaneously reproduce the observed equilibrium number of species and the qualitative trends in the RSA, SAR and SAF measures. While RSA and SAR are widely studied in the literature and a number of models focus on their reproduction, the spatial distribution of individuals of a given species draws less attention. For the BCI plot and for not too big distances, the SAF curves exhibit power-law decay. This is true for species with various abundance, ranging from a few hundreds of individuals up to the case of the most abundant species. The present study considers this measure equally important and besides the study of RSA and SAR, the spatial distribution of a given species (usually the most abundant species is chosen for better statistics) is also closely investigated.

Our previous attempt \cite{HorvatEtAl_2010} towards a more precise description of tropical tree communities failed in the reproduction of the spatial distribution of individual species by assuming constant probability of long-range interactions and favoring first neighbor interactions of individuals placed on a uniform grid. The present model comprises in a more realistic way the long-range interactions by assuming random diffusion of seeds. Also, the simulation method used for studying the model is improved by considering the coalescence technique which is faster and makes it possible to take into consideration the role of the surrounding metacommunity in determining the species diversity.

Using a speciation rate of $q=1.0 \times 10^{-5}$ and assuming a maximum survival distance of about 2500 m, the model produces results in agreement with the experimentally investigated ecological measures. The species-area scaling and species abundance distribution is reproduced in an acceptable manner, the autocorrelation function of the most abundant species shows the observed power-law decay in the limit of small distances as well. The strength of the model is that with only one freely adjustable parameter we are able to  qualitatively approach a quite complex phenomenon. Obviously, the model neglects many existing ecological processes, however, the results obtained so far suggest that this simple neutral approach captures many important aspects of tropical tree communities.

\section*{Acknowledgments}
Work supported by research grant PCCE 312/2008 (Complex Ideas).

\end{document}